# Carbon-fiber tips for scanning probe microscopes and molecular electronics experiments

**Gabino Rubio-Bollinger**[1,*], Andres Castellanos-Gomez[1,2,*], Stefan Bilan[3], Linda A. Zotti[3], Carlos R. Arroyo[1], Nicolás Agraït[1,4], Juan Carlos Cuevas[3]

[1] Departamento de Física de la Materia Condensada (C–III). Universidad Autónoma de Madrid, Campus de Cantoblanco, E-28049 Madrid, Spain.

[2] Kavli Institute of Nanoscience, Delft University of Technology, Post Office Box 5046, 2600 GA Delft, Netherlands.

[3] Departamento de Física Teórica de la Materia Condensada (C–III). Universidad Autónoma de Madrid, Campus de Cantoblanco, 28049 Madrid, Spain.

[4] Instituto Madrileño de Estudios Avanzados en Nanociencia IMDEA-Nanociencia, E-28049 Madrid, Spain.

gabino.rubio@uam.es, a.castellanosgomez@tudelft.nl

**ABSTRACT**

We fabricate and characterize carbon-fiber tips for their use in combined scanning tunneling and force microscopy based on piezoelectric quartz tuning fork force sensors. An electrochemical fabrication procedure to etch the tips is used to yield reproducible sub-100-nm apex. We also study electron transport through single-molecule junctions formed by a single octanethiol molecule bonded by the thiol anchoring group to a gold electrode and linked to a carbon tip by the methyl group. We observe the presence of conductance plateaus during the stretching of the molecular bridge, which is the signature of the formation of a molecular junction.

## BACKGROUND

Understanding electron transport through a single molecule is a basic goal in molecular electronics [1]. A primary goal is to find reliable ways to form a stable mechanical and electrical connection between the molecule and macroscopic electrodes. The mechanical and electrical properties of a molecular junction are not only determined by the molecular structure but also by the chemical nature of the electrodes [2].

Here, we have explored the use of carbon-based tips as contact electrodes to form molecular junctions [3]. Using the scanning tunneling microscope (STM) break-junction technique, we have measured the electrical conductance of several hundreds of octanethiol-based single-molecule bridges ($CH_3$-$C_7H_{14}$-SH) in which the thiol anchoring group is bound to a gold electrode and the methyl group is linked to a carbon electrode.

In order to form single-molecule junctions with a carbon electrode, we provide an STM with a carbon-fiber tip [4, 5]. The microscopic structure of the tip is composed by graphitic planes aligned parallel to the fiber longitudinal axis, yielding high electrical conductivity $\sigma = 7.7 \times 10^4$ S/m. Carbon-fiber tips are prepared from freshly cut individual carbon fibers obtained from a commercially available carbon fiber rope and are mounted in a home-built STM [6].

## METHODS:

Although the use of mechanically fabricated tips (by simply cutting a metallic wire) is rather common in STM, the atomic force microscope (AFM) resolution strongly depends on the tip sharpness because of long range interactions between the tip and the sample. We have developed an electrochemical procedure to etch carbon-fiber tips that yields sharp carbon-fiber tips.

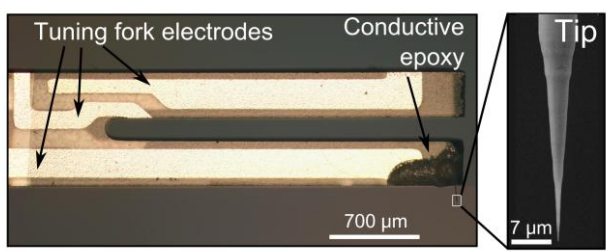

**Figure 1. Carbon-fiber tip mounted on a tuning fork.** (Left) Optical micrograph of a quartz tuning fork where the carbon-fiber tip is attached at the free end of one of its prongs. (Right) Scanning electron micrograph of an electrochemically etched carbon-fiber tip. The radius of curvature of the tip apex is 50 nm.

The setup used to electrochemically etch the carbon fibers [4] is similar to the one used to etch metallic tips [7]. A 5 to 10 mm long fiber is extracted from the fiber rope. One end of the fiber is immersed a few microns into a drop of 4 M KOH solution suspended in a 4-mm diameter gold ring. A voltage bias of 5 V is applied between the fiber end and the gold ring which is grounded. The etching takes place over a period of tens of seconds until the fiber breaks, opening the electrical circuit and stopping the etching. Afterwards, the fiber is rinsed with distilled water. Reproducible tips with sub-100-nm apex radius of curvature can be obtained following this procedure (Figure 1). The tip is then glued with conductive epoxy at the free end of one of the prongs of a miniature quartz tuning fork [8, 9]. The fiber is electrically connected to one of the tuning fork electrodes, which is grounded, in order to be able to simultaneously operate the microscope in STM mode [4].

**RESULTS AND DISCUSSION**

We have studied the interaction force between a carbon-fiber tip and a Au (111) surface when the tip is approached to the surface until reaching the electron tunneling regime. For simultaneous AFM operation, a frequency modulation mode has been used, driving the tuning fork at its resonance frequency using a phase-locked loop circuit [10].

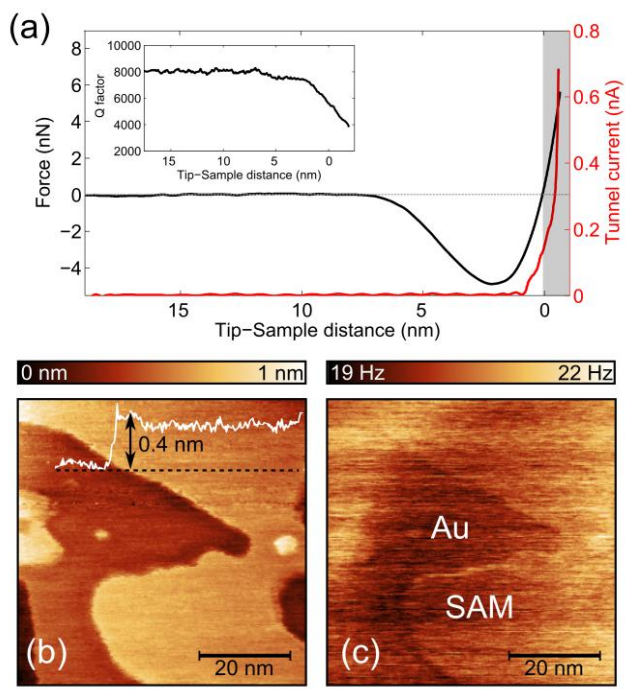

**Figure 2. Carbon-fiber tip and substrate interaction.** (**a**) Simultaneous measurement of force (black line, left axis) and tunnel current (red line, right axis) during a tip-sample approach. The inset shows the simultaneously measured quality factor of the tuning fork oscillation. (**b**) Topography image of an octanethiol self-assembled monolayer (SAM) on Au (111) sample obtained in the constant tunnel current STM mode and (**b**) the simultaneously measured frequency shift of the tuning fork oscillation which enables to unambiguously identify bare gold areas (dark) and SAM-covered regions (bright).

An attractive (repulsive) force gradient acting between tip and sample results in a positive (negative) shift of the resonance frequency. In the limit of a small oscillation amplitude (here 0.2 nm$_{RMS}$), the force gradient can be related to the frequency shift, and the force vs. distance curve (Figure 2a) can be obtained by integration [11]. The time-averaged tunnel current (Figure 2a) is simultaneously measured. We find that tunnel currents of up to 100 pA can be obtained while in the attractive force regime, that is, in the non-contact regime, indicating that the tip is not oxidized or contaminated. The tunnel current vs. tip-sample distance shows an exponential dependence corresponding to a tunnel barrier height of 0.8 eV, which is common for the environmental conditions of the experiment: room temperature in air [12]. We have also measured the change of the quality factor ($Q$) of the tuning fork oscillation during the approach (Figure 2a inset). The $Q$ factor falls by 40% before entering the tunnel regime and can be attributed to several sources such as ohmic dissipation or force gradient-induced imbalance of tuning fork prongs [13].

Single-molecule junctions are obtained by repeatedly forming and breaking the contact [2, 14, 15] between the tip and a gold substrate partially covered with an octanethiol self-assembled monolayer shown in Figure 2b. The molecules were deposited on a gold substrate (commercially available from Arrandee, Werther, Germany ) which was initially treated with piranha solution and then flame annealed to prepare a flat reconstructed Au (111) surface. The substrate was incubated for 12 h in a 1-mM solution of octanethiol (Sigma-Aldrich Corporation, MO, USA) in toluene followed by rinsing and sonicating in pure toluene and subsequently dried in a stream of helium gas. These deposition conditions are well known to yield a densely packed SAM [16]. Molecules are contacted by gentle repeated indentation of the tip into the substrate until the tunnel resistance is 2 MΩ. The tip is subsequently retracted, and the electrical conductance trace is measured. To overcome junction-to-junction fluctuations, we have performed a statistical analysis [17, 18] in which all junction realizations are taken into account to build a conductance histogram (Figure 3) from all the 640 conductance traces acquired at 20 different spots of the SAM. The hump in the histogram is associated with the presence of conductance plateaus in individual traces and the background with tunnel conduction. We find that the broad hump can be fitted to the sum of two Gaussian peaks in a linear conductance scale whose centers are located at $G_1 = (5.9 \pm 4.1) \times 10^{-6} G_0$ and $G_2 = (1.3 \pm 0.5) \times 10^{-5} G_0$, where $G_0$ is the conductance quantum ($2e^2/h$, with $e$ the electron electric charge and $h$ the Plank's constant). The fact that the value of $G_2$ is twice that of $G_1$ suggests that plateaus at conductance $G_2$ correspond to electron transport through two simultaneously connected molecules, each of which has a conductance $G_1$.

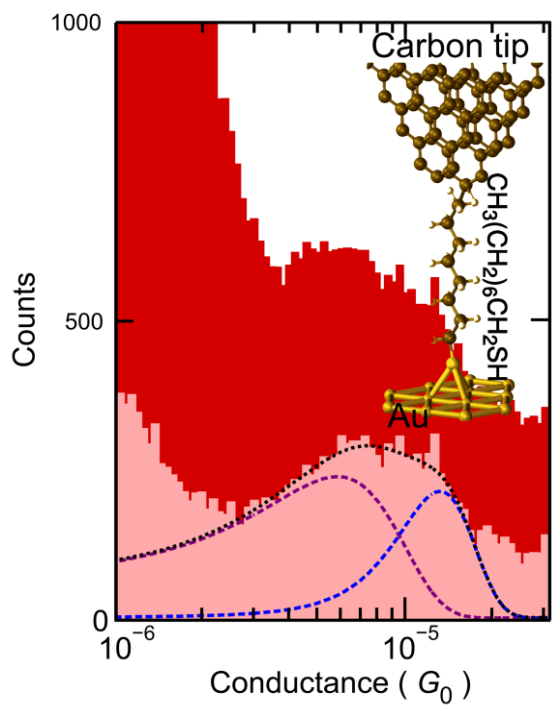

**Figure 3.** Octanethiol molecule conductance histogram. Conductance histogram built from 500 traces (dark red) and corrected histogram (light red) after subtracting the tunneling contribution. The corrected histogram shows two conductance peaks. The most probable conductance associated to these peaks is obtained from the fit of the corrected histogram to the sum of two Gaussian curves. The dashed purple and blue curves are the two Gaussians, and the dotted black curve is their sum. The positions of the maximum of the peaks mark the two most probable molecular configurations.

## CONCLUSION

We have fabricated and characterized carbon-fiber tips for their use in combined STM/AFMs based on quartz tuning fork force sensors. We develop an electrochemical procedure to etch carbon-fiber tips which yields sub-100-nm tip apex radius of curvature in a reproducible way, increasing the lateral resolution in AFM measurements. We show that carbon-fiber tips mounted on quartz tuning fork force sensors can be reliably used in force and/or tunnel current vs. distance measurements and simultaneous STM/AFM microscopy.

In addition, we have used carbon fiber tips as electrodes in an STM-break junction configuration to form single-molecule junctions with octanethiol molecules deposited on a gold surface. We find that carbon tips provide a stable mechanical linking to the methyl group allowing to form single-molecule bridges. Therefore, carbon tips can be suitable candidates to contact a variety of organic molecules, and they can also be combined with other substrate materials including carbon itself to form purely organic single-molecule devices..

## ABBREVIATIONS

AFM, atomic force microscope; SAM, self-assembled monolayer; STM, scanning tunneling microscope.

## ACKNOWLEDGMENT

This work was supported by MICINN (Spain) through the programs MAT2008-01735, MAT2011-25046 and CONSOLIDER-INGENIO-2010 'Nanociencia Molecular' CSD-2007-00010; Comunidad de Madrid through program Nanobiomagnet S2009/MAT-1726; and European Union through programs BIMORE (MRTN-CT-2006-035859) and ELFOS (FP7).